\documentclass[prx,aps,twocolumn,showpacs]{revtex4-1}

\usepackage{amssymb,amsmath}
\usepackage{graphicx}
\usepackage{multirow}
\usepackage{color}
\usepackage[english]{babel}
\usepackage{hyperref}
\hypersetup{
    colorlinks=true,
    linkcolor=blue,
    filecolor=blue,      
    urlcolor=blue,
    citecolor=blue
}
\usepackage{array}

\newcommand{\be}{\begin{equation}}
\newcommand{\ee}{\end{equation}}
\newcommand{\ba}{\begin{equation}}
\newcommand{\ea}{\end{equation}}
\newcommand{\bea}{\begin{eqnarray}}
\newcommand{\eea}{\end{eqnarray}}

\newcommand{\w}{\omega}

\newcommand{\tu}{\tilde{U}}

\newcommand{\te}{\tilde{\epsilon}}

\newcommand{\tz}{\tilde{z}}

\newcommand{\e}{\epsilon}
\newcommand{\Ai}{\mathrm{Ai}}

\newcommand{\eref}[1]{Eq.~(\ref{#1})}
\newcommand{\esref}[1]{Eqs.~(\ref{#1})}

\newcommand{\ocite}[1]{Ref.~\cite{#1}}

\begin{document}
\title{Theory of Silicon Spin Qubit Relaxation in a Synthetic Spin-Orbit Field}
\author{Amin Hosseinkhani}
\email{amin.hosseinkhani@uni-konstanz.de}
\thanks{current affiliation: IQM, Nymphenburgerstr. 86, 80636 Munich, Germany}
\affiliation{Department of Physics, University of Konstanz, D-78457 Konstanz, Germany}
\author{Guido Burkard}
\email{guido.burkard@uni-konstanz.de}
\affiliation{Department of Physics, University of Konstanz, D-78457 Konstanz, Germany}

\begin{abstract}
We develop the theory of single-electron silicon spin qubit relaxation in the presence of a magnetic field gradient. Such field gradients are routinely generated by on-chip micromagnets to allow for electrically controlled quantum gates on spin qubits. We build on a valley-dependent envelope function theory that enables the analysis of the electron wave function in a silicon quantum dot with an arbitrary roughness at the interface.  We assume the presence of single-layer atomic steps at a Si/SiGe interface, and study how the presence of a gradient field modifies the spin-mixing mechanisms. We show that our theoretical modeling can quantitatively reproduce results of experimental measurements of qubit relaxation in silicon in the presence of a micromagnet. We further study in detail how a  field gradient can modify the EDSR Rabi frequency of a silicon spin qubit.  While this strongly depends on the details of the interface roughness, interestingly, we find that adding a micromagnet on top of a spin qubit with an ideal interface can even reduce the EDSR frequency within some interval of the external magnetic field strength. 
\end{abstract}
	
\maketitle
\section{Introduction}
Silicon heterostructures have emerged as a very promising material platform for spin-based quantum information processing \cite{Zwanenburg13,GB21}. Recently two-qubit gates in silicon spin qubits were demonstrated with an overall fidelity exceeding $99\%$ by a number of experimental studies \cite{Mills21,Xue22,Noiri22,Madzik22}, a very important step towards realizing fault-tolerant silicon-based quantum computation. The intrinsic spin-orbit coupling (SOC) in silicon quantum dots is very weak and (largely) originates from the interface inversion asymmetry \cite{Golub04,Nestoklon06,Prada11}. While the weakness of the SOC is convenient as it limits the spin-mixing and thus the qubit relaxation rate, it also gives rise to slow electric dipole spin resonance (EDSR). The latter is a standard technique enabling electrical control of spin qubits. In order to perform faster EDSR, one possibility is to integrate a micromagnet in proximity to the quantum dot that generates a position-dependent magnetic field \cite{Yoneda18,Mi18-1}. This, in turn, gives rise to a synthetic SOC that influences the EDSR Rabi frequency as well as the qubit relaxation rate \cite{Borjans19}.

As we quantitatively show in this paper, the resulting additional spin-mixing due to the gradient field strongly depends on the roughness at the Si/barrier interface as well as the lateral size of the quantum dot. It is well known that in the presence of interface steps, the valley splitting energy $E_{\rm vs}$ is suppressed \cite{Friesen07,Boross16,Tariq,Hosseinkhani20}. Moreover, the presence of interface steps breaks the inversion symmetry within the interface plane and therefore, it generally gives rise to a non-vanishing in-plane dipole matrix elements. A recently developed valley-dependent envelope function theory based on the effective mass approximation  enables directly calculating the dipole matrix elements as well the spin-valley mixing caused by the intrinsic spin-orbit coupling \cite{Hosseinkhani21}. The theory also implies that the interface roughness can lead to strong anisotropic spin-valley mixing and spin relaxation. This anisotropic behavior has been experimentally found in both Si/SiGe  and CMOS quantum dots \cite{Zhang20,Zhang21,Spence21}.

Here we build on the valley-dependent envelope function of \ocite{Hosseinkhani21} and study in detail the influence of a gradient magnetic field on the relaxation rate and the EDSR Rabi frequency of a single-electron Si/SiGe spin qubit for various configurations of interface steps. For certain positions for the interface steps, we show that our theory can qualitatively explain the experimental data in \ocite{Borjans19} for the qubit relaxation time in the presence of the micromagnet with a minimal set of free parameters. The work presented here directly yields the dipole matrix elements, valley splitting energy, and the spin-valley coupling in the presence of the micromagnet and interface steps, which are quantities that have previously been treated by theory as free parameters.

The paper is structured as follows. In Sec.~\ref{sec:model} we establish our model for a single-electron spin qubit in Si in the presence of a micromagnet and interface disorder.  In Sec.~\ref{sec:dis} we discuss the effect of the micromagnet on the qubit relaxation rate and the EDSR Rabi frequency and consider a crossover from a disordered interface to a nearly flat interface, and show that our theory can reproduce the experimental data from \ocite{Borjans19}. In Sec.~\ref{sec:sum} we summarize the work and present our conclusions. The appendices contain further details of the analysis presented in the main part of the paper.

\section{Model}\label{sec:model}
\begin{figure}[t!]
\begin{center}
\includegraphics[width=0.45\textwidth]{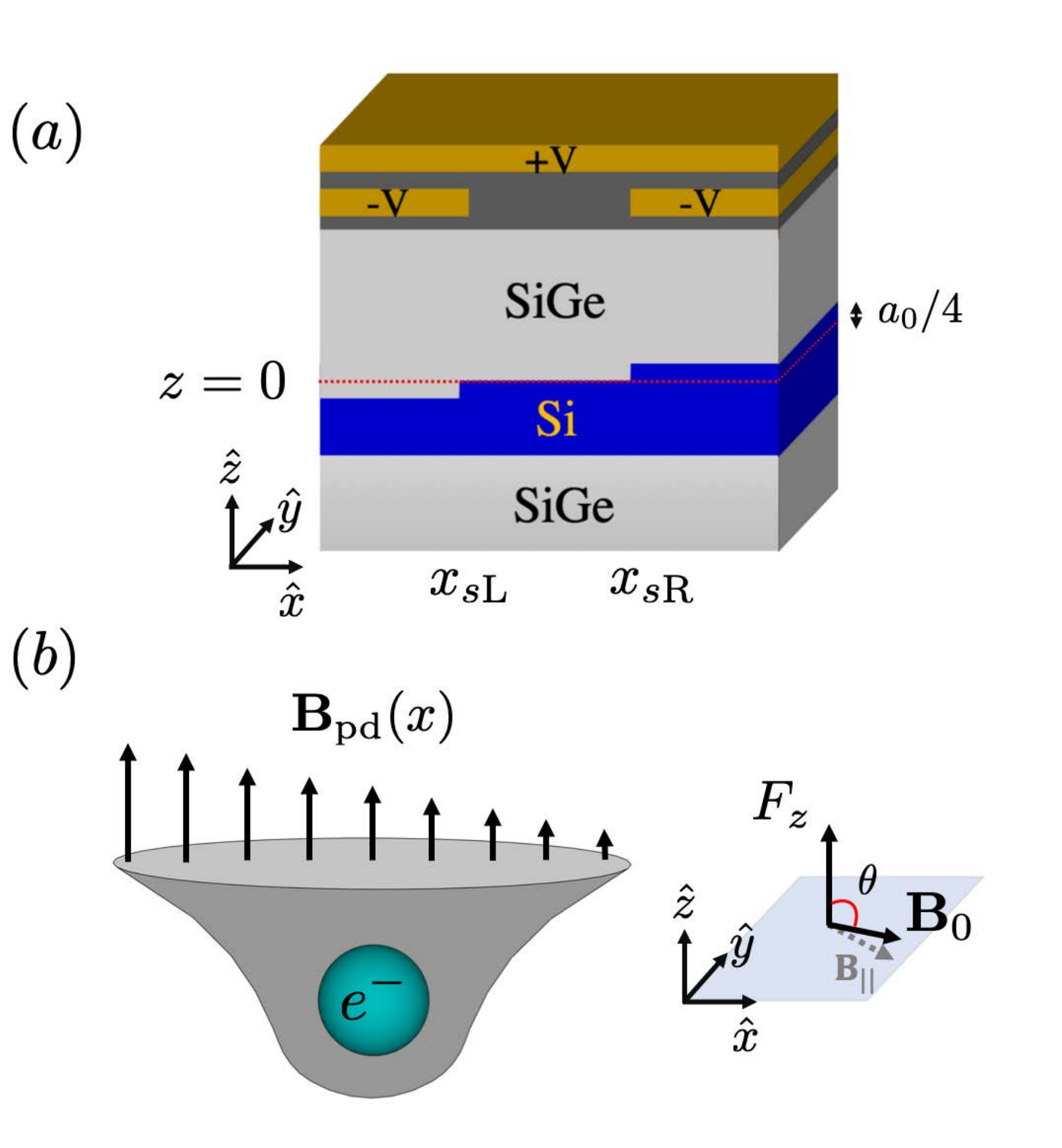}
\end{center}
\caption{(a) Schematic of a disordered quantum dot formed at a Si/SiGe interface comprising two atomic steps. The top gates with applied voltages $\pm V$ are used to trap and confine a single electron in the silicon layer. The atomic steps have the width $a_0/4$ where $a_0=0.543$ nm denotes the lattice constant. (b) Schematic of trapped single-electron in a confinement potential in the presence of a magnetic-field gradient, $\textbf{B}_{\rm pd}(x)$. In addition, there is an out-of-plane electric field $F_z$ and a homogeneous magnetic field $\textbf{B}_0$. The angle between $\textbf{B}_0$ and the $\hat{z}$ axis is denoted $\theta$.  }  \label{fig:qdot}
\end{figure}

 We consider a single-electron quantum dot inside a  SiGe/Si/SiGe heterostructure grown along the $\hat{z}$ direction ([001]). The potential offset between the minima of the conduction band in Si and SiGe amounts to $U_0=150$ meV.  An electric field is applied along $\hat{z}$ via the top gate electrodes. We assume that the lower SiGe/Si interface is ideally flat and located at $z=-d_t$. Taking the realistic value of $d_t=10$ nm, neglecting possible interface roughness at the lower interface is always justified by the fact that for $F_z \gtrsim 2$ MV/m, the amplitude of the wave function at the lower interface becomes negligible. Throughout this work, we set $F_z = 15$ MV/m. At the upper Si/SiGe interface, within our model and for simplicity, we allow for up to two single-layer interface steps located at the left and right side of the quantum dot center, $x_{x\mathrm{L}}$ and $x_{x\mathrm{R}}$, see Fig. \ref{fig:qdot}(a). 
 
 The electric gates surrounding the quantum dot give rise to an electrostatic potential leading to the confinement a single electron. In order to split up the spin states, an external in-plane magnetic field $\textbf{B}_{||}=(B_{||,x},B_{||,y},0)$ is applied. Moreover, we consider a magnetic field gradient, typically caused by a micromagnet, $\textbf{B}_{\rm mm}$. Following the experimental setup of \ocite{Borjans19}, we take $\textbf{B}_{\rm mm}=(b_x,0,b_z(x))$ in which $b_z(x)= b_{0z}+c_{\rm mm}x$. We therefore write the total magnetic field as a sum of homogeneous and position-dependent terms,
 \begin{align}
    \textbf{B}_t=\textbf{B}_0+\textbf{B}_{\rm pd}(x),
 \end{align}
in which $\textbf{B}_0=(B_{||,x}+b_x,B_{||,y},b_{0z})$ and $\textbf{B}_{\rm pd}=(0,0,c_{\rm mm}x)$, see Fig.~\ref{fig:qdot}(b). 
 
The spin qubit Hamiltonian then reads,
\begin{align}\label{eq:HS}
    H = H_c + H_z + H_{\mathrm{i-SOC}} + H_{\mathrm{s-SOC}},
\end{align}
where $H_c$ is dominant contribution to the total Hamiltonian and the other terms can be considered as a perturbation. This dominant term describes the quantum dot confinement in the presence of interface steps and magnetic field and it can be written as,
\begin{align} 
\label{eq:Hc}
H_c= H_0' + H_{||} + U_{\mathrm{steps}}(x,z).
\end{align}
Here $H_0'$ is the separable and exactly solvable Hamiltonian caused by the in-plane and out-of-plane confinements, and it is given by \eref{eq:H_xyz_B} in Appendix~\ref{app:1} (the prime index ($'$) used here indicates that the confinement frequencies are modified by the magnetic field.) $H_{||}$ is due to the the couplings induced by the total in-plane magnetic field and it is given by \eref{eq:H_||}. $U_{\mathrm{steps}}(x,z)$ is a change to the offset potential of an ideally flat interface due to the presence of interface steps and within our model it reads (note that the offset potential of an ideal interface is included in $H_0'$),
\begin{align}
\label{eq:U_pert}
U_{\mathrm{steps}}(x,z)&=U_0\theta(-z)\theta\left(z+\frac{a_0}{4}\right)\theta(x_{s\mathrm{L}}-x) \nonumber\\
  &-U_0\theta(z)\theta\left(z-\frac{a_0}{4}\right)\theta(x-x_{s\mathrm{R}}) .
\end{align}

The valley-dependent envelope functions $\Psi_{xyz}^{j}$ can be determined from the equation
\begin{align}
\label{eq:VDenv}
\sum_{j=\pm z}a_je^{ik_jz}\left\{H_c+V_v(r)-E\right\}\Psi_{xyz}^{j}=0,
\end{align}
in which $V_v(r)=V_v\mathcal{S}_\mathrm{int}(x,z)$ is the valley coupling that vanishes everywhere except at the Si/barrier interface \cite{Friesen10}. The quantity $V_v$ has been modeled in \ocite{Hosseinkhani21} as a function of the offset potential $U_0$, the electric field $F_z$ and details of the periodic parts of the Bloch functions. We show the form of $V_v$ in \eref{eq:Vv} in Appendix~\ref{app:1} for completeness. Within our model for the disordered interface, the interface function reads,
\begin{align}\label{eq:Sint}
    \mathcal{S}_\mathrm{int}(x,z)&=\delta\left(z+\frac{a_0}{4}\right)\theta(x_{s\mathrm{L}}-x)\nonumber\\
    &+\delta(z)\theta(x-x_{s\mathrm{L}})\theta(x_{s\mathrm{R}}-x)\nonumber \\
&+\delta\left(z-\frac{a_0}{4}\right)\theta(x-x_{s\mathrm{R}}).
\end{align}
It has been discussed in detail in \ocite{Hosseinkhani21} how to find the valley-dependent envelope functions by solving \eref{eq:VDenv}; we review the solutions in Appendix \ref{app:2}.   

 The term $H_z=\frac{1}{2}g\mu_B\boldsymbol{\sigma}\cdot\textbf{B}_0$ in \eref{eq:HS} is the Zeeman splitting caused by the homogeneous magnetic field, while $H_{\mathrm{i-SOC}}$ in the same equation describes the intrinsic spin-orbit coupling in the silicon quantum well that is caused by the interface inversion asymmetry and contains Rashba and Dresselhaus-like terms. For a disordered quantum dot $H_{\mathrm{i-SOC}}$ can be written as \cite{Hosseinkhani21},
 \begin{align}
     H_{\mathrm{i-SOC}}=H_R+H_D,
 \end{align}
 in which,
 \begin{align}\label{eq:H_R}
    H_R&=\frac{1}{2}\gamma_R\left\{ p_y\sigma_x-p_x\sigma_y,\mathcal{S}_\mathrm{int}(x,z)\right\},\\\label{eq:H_D}
    H_D&=\frac{1}{2}\gamma_D\cos\left(\frac{4\pi z}{a_0}\right)\left\{p_x\sigma_x-p_y\sigma_y,\mathcal{S}_\mathrm{int}(x,z)\right\}.
\end{align}
Note that the factor of $\cos(4\pi z/a_0)$ in the Dresselhaus term is due to the fact that the coefficient of the Dresselhaus spin-orbit interaction changes sign when encountering a single-layer atomic step at the interface \cite{Ferdous18,FerdousPRB18,Jock18}. Finally, the last term in \eref{eq:HS} is the synthetic spin-orbit coupling which originates from the Zeeman splitting due to position-dependent part of the magnetic field,
\begin{align}
    H_{\mathrm{s-SOC}}=\frac{1}{2}g\mu_B\boldsymbol{\sigma}\cdot\textbf{B}_{pd}(x).
\end{align}
\subsection{Modified qubit levels due to spin-valley mixing}
Having reviewed our model for a single-electron spin qubit in the presence of interface steps and magnetic field-gradient, we now turn to study the modified qubit levels. We begin by considering a case where the orbital splitting is the dominant energy scale so that we can neglect the higher orbital states. In this case, the qubit levels are modified due to spin-valley mixing (SVM). In the absence of the spin-orbit coupling, we can then only consider the unperturbed states, 
\begin{align}
\label{eq:4b}
  |1\rangle &= |\nu^{(q=0)},\downarrow\rangle ,\quad |2\rangle = |\nu^{(q=0)},\uparrow\rangle , \nonumber \\
  |3\rangle &= |\nu^{(q=1)},\downarrow\rangle ,\quad |4\rangle = |\nu^{(q=1)},\uparrow\rangle,
\end{align}
in which the ground ($q=0$) and excited ($q=1$) valley-orbital states read,
\begin{align}\label{eq:v0}
    |\nu^{(q=0)}\rangle=\frac{1}{\sqrt{2}}\left\{|+z^{(0)}\rangle-e^{-i\phi_v}|-z^{(0)}\rangle\right\},\\\label{eq:v1}
    |\nu^{(q=1)}\rangle=\frac{1}{\sqrt{2}}\left\{|+z^{(1)}\rangle+e^{-i\phi_v}|-z^{(1)}\rangle\right\},
\end{align}
where $\phi_v$ is the valley phase and
\begin{align}\label{eq:pmZ}
|\pm z^{(q)}\rangle=&e^{\pm ik_0z}u_{\pm z}(r)\Psi_{xyz}^{\pm z,(q)},
\end{align}
where $u_{\pm z}(r)$ are the periodic part of the wave function and $\Psi_{xyz}^{\pm z,(q)}$ is the valley-dependent envelope function that can be found by solving \eref{eq:VDenv}, see Appendix ~\ref{app:2}. 

The spin-orbit terms, $H_{\mathrm{i-SOC}}$ and $H_{\mathrm{s-SOC}}$, give rise to the coupling between the unperturbed states and mix the spin states.
In general we can write for the matrix elements of the spin-valley coupling,
\begin{align}
  \Delta_{ij}=\langle i|H_{\mathrm{s-SOC}}+H_{\mathrm{i-SOC}}|j\rangle  =\Delta_{ij}^\mathrm{s-SOC}+\Delta_{ij}^\mathrm{i-SOC}.
\end{align}
The matrix elements of the intrinsic spin-orbit interaction $\Delta_{ij}^\mathrm{i-SOC}$ can be readily found using \esref{eq:H_R} and (\ref{eq:H_D}) as well as the valley-dependent envelope function \eref{eq:Psi_t_p}, as discussed in detail in \ocite{Hosseinkhani21}. Here we take the matrix elements of $\Delta_{ij}^\mathrm{i-SOC}$ as given quantities, and focus on the matrix elements of the synthetic spin-orbit interaction. By defining $c_c=\frac{1}{2}g\mu_Bc_{mm}$ and $x_{ij}=\langle\nu^{(q=i)}|x|\nu^{(q=j)}\rangle$,
we find
\begin{align}
\label{eq:D_ssoc}
    \Delta_{21}^\mathrm{s-SOC}&=c_cx_{00}\sigma_z^{\uparrow\downarrow},\quad  \Delta_{32}^\mathrm{s-SOC}=c_cx_{10}\sigma_z^{\downarrow\uparrow}, \nonumber\\
    \Delta_{41}^\mathrm{s-SOC}&=c_cx_{10}\sigma_z^{\uparrow\downarrow},\quad     \Delta_{43}^\mathrm{s-SOC}=c_cx_{11}\sigma_z^{\uparrow\downarrow} .
\end{align}

Note that the Pauli matrix $\sigma_z$ is defined with respect to the lattice crystallographic axes whereas the spin states are defined with respect to the direction of  $\textbf{B}_0$. It is easy to verify that $   \sigma_z^{\uparrow\downarrow}=\sigma_z^{\downarrow\uparrow}=-\sin\theta$ in which $\theta$ is the angle between $\textbf{B}_0$ and the $\hat{z}$ axis, as shown in Fig.~\ref{fig:qdot}(b). Furthermore, as we see from \eref{eq:D_ssoc}, the matrix elements of the synthetic spin-orbit interaction are proportional to the dipole matrix elements which, in turn, strongly depend on the interface roughness. Indeed, for an ideally flat interface, all in-plane dipole moments vanish due to the mirror symmetry. In that case, \eref{eq:D_ssoc} reveals that the presence of the micromagnet does not affect the qubit levels via the spin-valley mixing (however, as we show in Sec.~\ref{sec:2b}, the qubit levels in this case are still changed by the presence of the micromagnet due to the spin-orbit mixing). The valley-dependent envelope function theory developed in \ocite{Hosseinkhani21} enables us to directly calculate all dipole moments shown in \eref{eq:D_ssoc} for an arbitrary configuration of interface steps. 

In the absence of intravalley spin-valley couplings, $\Delta_{21}, \Delta_{43}=0$, it is easy to find an analytical relation for the modified qubit levels, as explicitly shown in Refs. \cite{Hosseinkhani21,Huang21}, by simply diagonalizing a $2\times 2$ matrix. As we consider the presence of the intravalley spin-valley couplings, here we numerically diagonalize a $4\times 4$ matrix enabling us to find the modified qubit ground and exited states, $|\tilde{g}\rangle$ and $|\tilde{e}\rangle$, due to the SVM. We note here that above the spin-valley hotspot, there is a physical state, $|\tilde{d}\rangle$,  between the qubit ground and excited states \cite{Hosseinkhani21,Huang21} and we also consider this state when finding the relaxation rate in the next section.

\subsection{Modified qubit levels due to spin-orbit mixing}\label{sec:2b}
We now consider the effects that arise due to the coupling to the higher orbital states. As already shown before in Refs. \cite{Yang2013,Huang14}, in a disordered quantum dot the spin-orbit mixing (SOM) becomes the dominant mixing mechanism at high magnetic fields above the spin-valley hotspot. In this case, for simplicity we neglect the excited valley state and find the modified qubit levels due to the SOM within our model read,
\begin{align}\label{eq:ql_SOM_g}
    |\tilde{g}\rangle &\simeq |0,\downarrow\rangle + c_1'|1_x,\uparrow\rangle + c_2^\mathrm{i-SOC}|1_y,\uparrow\rangle,\\\label{eq:ql_SOM_e}
    |\tilde{e}\rangle &\simeq |0,\uparrow\rangle + c_3'|1_x,\downarrow\rangle + c_4^\mathrm{i-SOC}|1_y,\downarrow\rangle,
\end{align}
in which $c_{1(3)}'=c_{1(3)}^\mathrm{i-SOC}+c_{1(3)}^\mathrm{s-SOC}$.   The contributions that are due to the intrinsic spin-orbit coupling are explicitly given in \ocite{Hosseinkhani21}. Here we find the additional terms that are due to the synthetic spin-orbit coupling read, 
\begin{align}
    c_1^\mathrm{s-SOC}= -\frac{1}{\sqrt{2}}\frac{c_cx_0'}{E_z+\hbar\omega_x'}\sigma_z^{\uparrow\downarrow},
\end{align}
and $c_3^\mathrm{s-SOC}$ is found from the above equation by replacing $E_z\rightarrow -E_z$ (note that $\sigma_z^{\uparrow\downarrow}=\sigma_z^{\downarrow\uparrow}$.) Here $E_z$ is the Zeeman splitting and $\hbar\omega_x'=\hbar^2/m_tx_0'^2$ is the in-plane orbital splitting modified by the magnetic field, see \eref{eq:om}.

\subsection{Relaxation rate and EDSR Rabi frequency}
\label{sec:2c}
Given the modified qubit levels in the presence of SVM and SOM that we presented earlier in this section, we now review the qubit relaxation rate and the EDSR Rabi frequency. The relaxation rate due to any source of an electric noise can be written as,
\begin{align}\label{eq:G_fi}
\frac{1}{T_1}=\frac{4\pi e^2}{\hbar^2} S_E(\omega)\sum_{j}|\langle \tilde{e}|r_j|\tilde{g}\rangle|^2,
\end{align}
in which the forms of the excited, $|\tilde{e}\rangle$, and ground, $|\tilde{g}\rangle$, qubit states depend on whether we consider SOM or SVM. Here $S_E(\omega)$ is the electric noise power due to combination of electron-phonon interaction, Johnson noise and $1/f$ charge noise,
evaluated at the qubit frequency $\omega=(E_{\tilde{e}}-E_{\tilde{g}})/\hbar$. The electron-phonon interaction is studied in several other publications for Si quantum dots \cite{Hosseinkhani21,Boross16,Hollmann20,Yang2013,Huang14} (as well as for GaAs quantum dots in \ocite{Khaetskii01,Golovach04}). The corresponding electric noise power in silicon is given by, e.g., Ref. \cite{Hollmann20}. The electric noise power of the Johnson noise due to a lossy transmission line is also given by \ocite{Hollmann20}, and we consider the general form for the $1/f$ noise power,
\begin{align}
    S^{1/f}_E(\omega)=\frac{S_0}{\omega^\alpha},
\end{align}
where $S_0$ determines the power spectral density at 1 Hz and the exponent $\alpha$ is device-dependent and it is typically reported to be between 0.5 and 2 for silicon quantum dots \cite{Kranz20}. We note that the fluctuations in the electric field can be mapped into the voltage fluctuations by the phenomenological relation,
\begin{align}
    S_E(\omega)=\frac{S_V(\omega)}{(el_0)^2}
\end{align}
where $l_0$ is a phenomenological length describing the distance between the spin qubit and the trapped fluctuating two-level system \cite{Hollmann20}. Very often the $1/f$ voltage noise power at 1 Hz is measured in the experiment \cite{Borjans19}, and we use the above relation to connect it to the electric noise power.

One promising way to electrically manipulate the states of spin qubits is via electric dipole spin resonance (EDSR). This has been performed in GaAs  and Si \cite{Yoneda18} spin qubits. In both cases, the spin manipulation is made possible by applying an ac electric field $E_\mathrm{ac}\cos(\omega t)$, which, at the leading order, enables controlled qubit transition via electric dipole moment. For GaAs, the dipole moment between the qubit states is caused due the SOM. For Si spin qubits, in addition to the SOM, the dipole moment can be caused by the SVM in the presence of interface roughness. 

In general the EDSR Rabi frequency can be written as
\begin{align}\label{eq:omedsr}
    \Omega_R=eE_\mathrm{ac}|\langle \tilde{g}|\textbf{r}|\tilde{e}\rangle|/\hbar
\end{align}
in which, as discussed earlier in this section, the modified qubit ground $|\tilde{g}\rangle$ and excited $|\tilde{e}\rangle$ states depend on whether SVM or SOM are the dominant mixing mechanism, and $\textbf{r}=(x,y,z)$. In either case, the presence of a micromagnet leads to additional terms to the spin-orbit interaction and therefore can give rise to a faster EDSR.  We stress again here that when the Si/barrier interface is ideally flat, the in-plane dipole matrix elements due to SVM vanish. As such, placing a micromagnet on top of a quantum dot with an ideally flat interface does not affect SVM-induced EDSR whereas it still affects the SOM-induced EDSR as we show in the following section. 

Through this work, we neglect potential interference between SVM and SOM contributions so that both the relaxation rate and the EDSR Rabi frequency are taken equal to contributions that separately originate from SVM and SOM mixing. This simplification is justified due to the fact that, except within a narrow interval for the magnetic field, either SVM or SOM give rise to the dominant contribution to the qubit levels so that possible interference between them is negligible.  In the next section we present our results on qubit relaxation and modified ESDR in the presence of a micromagnet.  

\section{Discussion}\label{sec:dis}
\begin{figure}[t!]
\begin{center}
\includegraphics[width=0.45\textwidth]{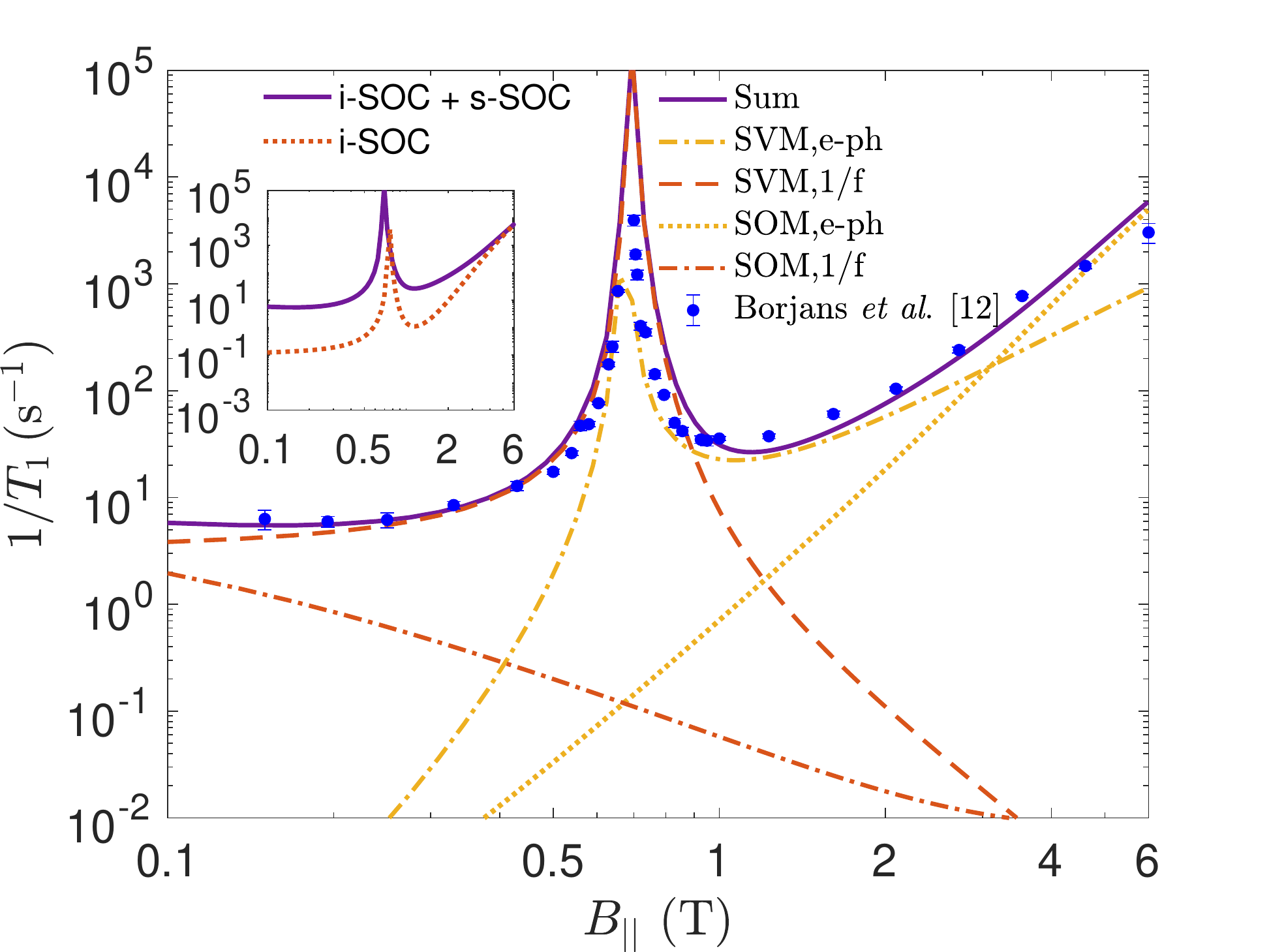}
\end{center}
\caption{Qubit relaxation rate $1/T_1$ as a function of external in-plane magnetic field $B_{||}$ in the presence of a micromagnet. Here the step positions are assumed to be at $x_{s\mathrm{L}}=- 0.9x_0$, $x_{s\mathrm{R}}=0.275 x_0$.
The violet solid curve describes the total calculated spin relaxation rate $1/T_1$, whereas the other curves indicate the contributions due to electron-phonon interaction (yellow dot-dahsed and dotted) and the $1/f$ charge noise (red dashed and dot-dashed). Experimental data points from \ocite{Borjans19} are shown as blue circles. Following \ocite{Borjans19}, we set $\phi_B=\pi/4$, $T_\mathrm{el}=115$ mK, $c_{mm}=1.8$ T/$\mu$m, $S_0$ =3$\mu$eV/$\sqrt{\mathrm{Hz}}$. We also set $b_x=92$ mT and $b_{0z}=15$ mT. Inset: Comparison of the relaxation rate in the presence (violet solid curve) and in the absence (red dotted) of the micromagnet.}  \label{fig:T1_fit}
\end{figure}

In Fig.~\ref{fig:T1_fit} we show our theoretical prediction for the qubit relaxation rate $1/T_1$ for a specific configuration of interface steps and compare our findings with experimental measurements from \ocite{Borjans19}. In order to arrive at our theoretical prediction, we first searched for a set of positions for the interface steps that gives rise to the same valley splitting as observed in the experiment ($E_{vs}\simeq 80.4\,\mu$eV). Among a number of possibilities, we find that setting $x_{s\mathrm{L}}=-0.9x_0$ and  $x_{s\mathrm{R}}=0.275x_0$ leads to the best fit to the experimental data. At the next step, we consider high magnetic fields above the spin-valley hotspot where, within our model, the relaxation rate is determined by the in-plane orbital splitting along $\hat{x}$ direction. We find that setting $\hbar\omega_x=3.1\mu$eV can fit the experimental data (we note here that in the original publication \cite{Borjans19} the in-plane orbital splitting is reported to be 2 $\mu$eV. This discrepancy could be an indication that the quantum dot in the experiment is elliptical).  

We then also consider the range of low magnetic fields below the spin-valley hotspot. Similar to \ocite{Hosseinkhani21} here we found that considering only the Johnson noise could not explain the low B-field behaviour of the relaxation rate. The amplitude of voltage noise at 1 Hz is reported from the experiment to be $3\mu$eV$/\sqrt{\mathrm{Hz}}$ \cite{Borjans19}. Having  fixed this, the fit shown in Fig.~\ref{fig:T1_fit} is achieved by setting $l_0=26.3$ nm and $\alpha=1.9$. For the coefficients of the intrinsic spin-orbit coupling, we use the same values found in \ocite{Hosseinkhani21}
 by fitting the theory to the experimental data for $1/T_1$ in the absence of a micromagnet. 
 
With this set of parameters for a disordered quantum dot, in the inset of Fig.~\ref{fig:T1_fit} we compare the relaxation rate $1/T_1$ in the presence of the micromagnet (so that the spin-orbit interaction is due to both intrinsic and synthetic terms) with the case in the absence of the micromagnet (so that only the intrinsic spin-orbit interaction is present). We observe that in the presence of the micromegnet, due to the additional spin-orbit mixing caused by the synthetic spin-orbit interaction, the relaxation rate is significantly increased particularly at low magnetic fields at which SVM determines the modified qubit levels. We also note here that for the SVM-induced contribution to the relaxation rate, there is an additional decay channel $|\tilde{e}\rangle\rightarrow |\tilde{d}\rangle$ above the spin-valley hotspot that we also take into account in finding all the results presented in Fig.~\ref{fig:T1_fit}.

\begin{figure}[t!]
\begin{center}
\includegraphics[width=0.45\textwidth]{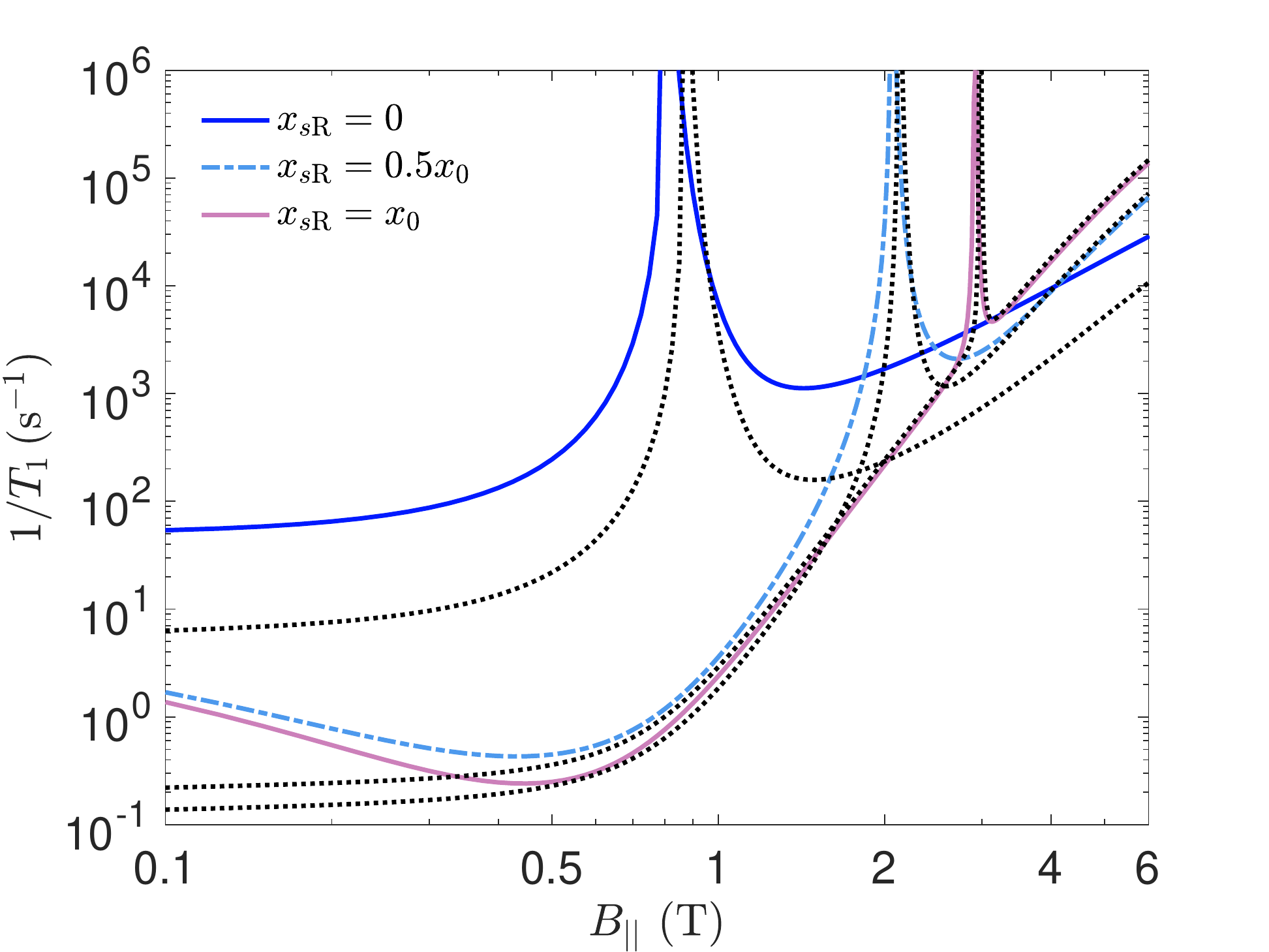}
\end{center}
\caption{The qubit relaxation rate $1/T_1$ as a function of the external in-plane magnetic field $B_{||}$ for various positions of a single interface step, $x_{s\mathrm{R}}$. The solid lines are obtained in the presence of the magnetic field and the black dotted lines are obtained in the absence of the micromagnet. All the other parameters are the same as given by the caption of Fig.~\ref{fig:T1_fit}.  }  \label{fig:T1_xR}
\end{figure}

As discussed in Sec.~\ref{sec:2b}, the spin-valley mixing caused by the synthetic spin-orbit interaction is proportional to the intravalley and intervalley dipole moments, see \eref{eq:D_ssoc}. These dipole matrix elements strongly depend on the interface roughness \cite{Hosseinkhani20}, and they vanish in an ideally flat interface due to the mirror symmetry. In Fig.~\ref{fig:T1_xR}, we show the qubit relaxation rate in the presence and in the absence of the micromagnet for a number of different configurations for the disordered interface. For simplicity, here we assumed there is only one single interface step located at $x_{s\mathrm{R}}$. At $x_{s\mathrm{R}}=0$, we observe that the behaviour of the relaxation rate is qualitatively similar to what is shown in Fig.~\ref{fig:T1_fit}, and within the range of the magnetic fields shown in Fig.~\ref{fig:T1_xR}, the relaxation rate increases in the presence of the micromagnet. 

However, noticeably for $x_{s\mathrm{R}}\gtrsim0.5x_0$, where the quantum dot approaches being flat, we observe a nonmonotonic behavior for the relaxation rate below the spin-valley hotspot in the presence of the micromagnet. This happens due to the fact that the dipole moments introduced in \esref{eq:D_ssoc} quickly decay when the single interface step is located further away from the quantum dot center so that in this case the spin mixing due to SVM becomes very small. On the other hand, the presence of the micromagnet substantially enhances the spin mixing due to SOM at low magnetic fields. As such, there is a competition between contributions of SOM,$1/f$ and SVM,$1/f$ to the total relaxation rate. Indeed, the qubit decay rate in the presence of the micromagnet at the magnetic fields below the minimum of $1/T_1$ is dominated by SOM and $1/f$ charge noise which decreases by increasing the magnetic field. 
At higher fields closer to the hotspot, the SVM finally becomes the dominant mixing mechanism giving rise to the observed nonmonotonic behavior.

We also note that in the limit where the quantum dot is nearly flat ($x_{sR}\gtrsim x_0$), the presence of the micromegnet hardly changes the behavior of the relaxation rate for magnetic fields above the minima of the relaxation rate. This is due to the fact that, as the single interface step is moved away from the dot center, the effective (spatially averaged) strength of the intrinsic spin-orbit interaction is greatly enhanced due to the behaviour of the Dresselhaus term, see \eref{eq:H_D}, so that, within the realistic parameters considered here for the micromagnet, the SOM becomes dominated by the intrinsic spin-orbit interaction. We finally note note here that Fig.~\ref{fig:T1_xR} shows that at high magnetic fields the relaxation rate at $x_{sR}=0.5x_0$ and $x_{sR}=x_0$ is slightly larger in the absence of the micromagnet. This is attributed to the interference between i-SOC and s-SOC contributions to the SOM as one can realize for the coefficients $c_1'$ and $c_3'$ introduced by \esref{eq:ql_SOM_g} and (\ref{eq:ql_SOM_e}).

\begin{figure}[t!]
\begin{center}
\includegraphics[width=0.45\textwidth]{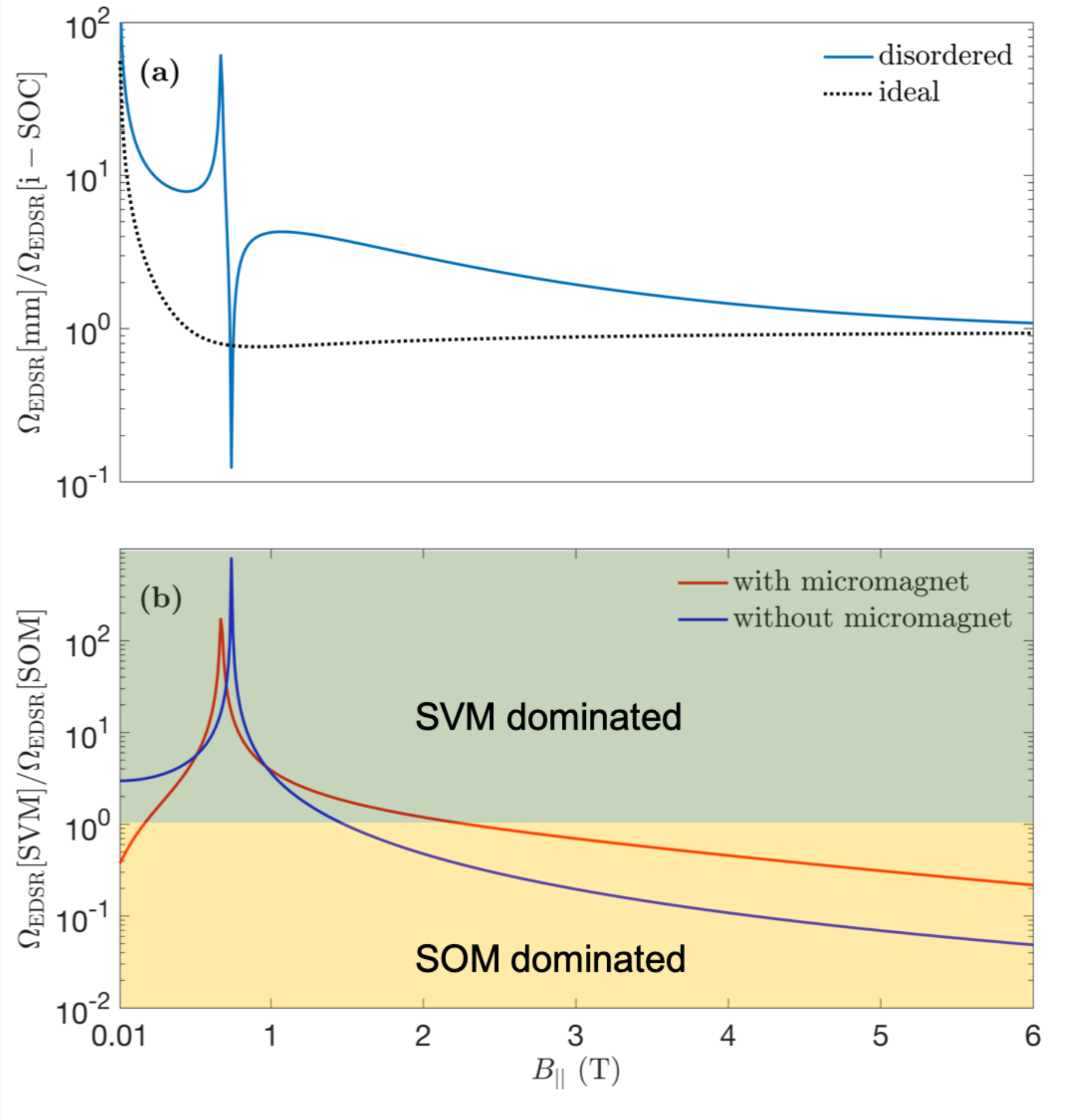}
\end{center}
\caption{ (a) EDSR Rabi frequency in the presence of the micromagnet relative to the EDSR frequency in the absence of the micromagnet as a function of the external magnetic field for both disordered and ideally flat quantum dot. (b) The SVM-induced contribution to the EDSR frequency relative to the SOM-induced contribution to the EDSR frequency for a disordered quantum dot. All quantum dot parameters are the same as given by the caption of Fig.~\ref{fig:T1_fit}. }  \label{fig:edsr}
\end{figure}

We now turn to study how the presence of the micromagnet can enhance the EDSR Rabi frequency. In Fig.~\ref{fig:edsr}(a), we consider a quantum dot with disorder interface (with steps' locations the same as in Fig.~\ref{fig:T1_fit}) as well as with an ideally flat interface and show the EDSR Rabi frequency in the presence of the micromagnet normalized to the EDSR frequency in the absence of the micromagnet as a function of the external in-plane magnetic field. For the disordered quantum dot, the rapid change of the behavior of the plot at $B_{||}\sim 0.7$ T is due to the fact that the spin-valley hotspot, at which the dipole moment between the qubit states given in \eref{eq:omedsr}, happens at slightly smaller external in-plane magnet field $B_{||}$ in the presence of the micromagnet, see the definition of $B_0$ in Sec.~\ref{sec:model}. We observe that adding a micromagnet can substantially enhance the EDSR Rabi frequency for a disordered quantum. For an ideally flat interface, on the other hand, adding a micromagnet can only enhance the EDSR frequency within some narrow interval at low magnetic fields. 

In the absence of a micromagnet, the dipole moment between the qubit states tends to vanish as the magnetic field goes to zero due to the time-reversal symmetry. As such, adding the micromagnet, that breaks the time-reversal symmetry, greatly enhance the EDSR frequency at low magnetic fields. As the magnetic field is increased for a quantum dot with an ideally flat interface, the intrinsic spin-orbit coupling quickly becomes the dominant mixing mechanism (within the realistic parameters for the micromagnet considered here) so that the synthetic spin-orbit coupling only plays a minor role in modifying the EDSR frequency. We note that within some interval of $B_{||}$, we observe that EDSR frequency is in fact slower in the presence of a micromagnet for an ideally flat interface. This behavior is again due to the interference between i-SOC and s-SOC contributions to the SOM that was mentioned earlier. 

In Fig.~\ref{fig:edsr}(b) we consider the same disordered quantum dot as in Fig.~\ref{fig:T1_fit} and study how the SVM and SOM contributions to the total EDSR frequency compare for the cases in the presence and in the absence of the micromagnet. While it is generally expected that at high magnetic fields for a disordered quantum dot, the SOM becomes the dominant mixing mechanism, we observe that this happens at a higher $B_{||}$ in the presence of the micromagnet. Furthermore,  we observe that within some narrow interval at low magnetic fields $B_{||} \lesssim 0.16$ T the dominant mixing mechanism in the presence of the micromagnet is the SOM.

\section{Summary and Outlook}\label{sec:sum}
Silicon-based heterostructures and quantum dots are a very promising platform for quantum information processing. While the relatively weak spin-orbit interaction enables long relaxation times, exceeding one second at low magnetic fields \cite{Borjans19}, it also gives rise to (relatively) slow qubit gates. A typical technique to achieve faster qubit gates is to fabricate a micromagnet in proximity to the quantum dot. The magnetic field gradient due to the micromagnet gives rise to a synthetic spin-orbit interaction that can boost the EDSR Rabi frequency.  In this work we studied in detail how the presence of a micromagnet can modify the qubit levels in a single-electron silicon spin qubit in Si/SiGe quantum dot. Finding the modified qubit levels is indeed the key to the quantitative analysis of the behavior of the spin relaxation and EDSR Rabi frequency, and we showed that the roughness at the Si/barrier interface is a crucial parameter that determines the influence of the micromagnet to the qubit levels. 

We build on the valley-dependent envelope function theory from our earlier work \ocite{Hosseinkhani21} that enables us to find the electron wavefunction in a quantum dot with an arbitrary interface roughness. In Sec.~\ref{sec:model} we summarize the essential aspects of the valley-dependent envelope function theory in the presence of interface steps and present how we can model the synthetic spin-orbit interaction. We also discuss the modified qubit levels caused by the micromagnet due to both spin-valley mixing and spin-orbit mixing. We find that the interface roughness strongly affects how a micromagnet can alter the qubit levels due to the SVM. Indeed, for an ideal quantum dot and as long as SVM is concerned, the micromagnet does not change the qubit levels at all. However, the modifications due to the SOM is (at the leading order) independent on the interface roughness and it only depends on the quantum dot lateral size (that determines the in-plane orbital splitting). In Sec.~\ref{sec:2c} we have reviewed the qubit relaxation rate and EDSR Rabi frequency. 
Finally, in Sec.~\ref{sec:dis} we have presented and discussed our results on the spin qubit relaxation rate and EDSR Rabi frequency. We showed that our modeling can quantitatively reproduce and explain experimental measurements for the  qubit relaxation time in the presence of a micromagnet for all ranges of magnetic field.

Building on the valley-dependent envelope function theory used here, future work can also analyze the effect of electric noise on the phase coherence time $T_2$ of spin qubits in silicon in the presence of a micromagnet-generated magnetic field gradient.

\appendix
\section{Quantum dot confinement Hamiltonian}\label{app:1}
In \eref{eq:Hc} we show the general form of the quantum dot confinement Hamiltonian in the presence of interface steps and an in-plane magnetic field. The contribution $H_0'$ in \eref{eq:Hc} reads,
\begin{align}
\label{eq:H_xyz_B}
H_{0}'=&\frac{p_x^2}{2m_t}+\frac{1}{2}m_t\w_x'^2x^2 +\frac{p_y^2}{2m_t}+\frac{1}{2}m_t\w_y'^2y^2\nonumber\\&+\frac{p_z^2}{2m_l}-eF_zz+ U(z),
\end{align}
in which $m_t=0.19\:m_e$ and $m_l=0.98\:m_e$ are the transverse and longitudinal effective mass, and the out-of-plane potential profile for a SiGe/Si/SiGe reads 
\begin{align}
\label{eq:Uz}
U(z)=U_0\theta(-z-d_t) + U_0\theta(z) + U_{\infty}\theta(z-d_b),
\end{align}
where $U_0=150$ meV is the energy offset between the minima of the conduction band in Si and SiGe,  $d_t$ is the thickness of the silicon layer (located between $-d_t\leq z\leq 0$) and $d_b$ is the thickness of the upper SiGe barrier. 

In the absence of a magnetic field, we can write for the confinement frequencies $\w_{x}=\hbar/m_tx_0^2$ and $\w_{y}=\hbar/m_ty_0^2$ in which $x_0(y_0)$ is the radious of the quantum dot along $\hat{x}(\hat{y})$. The presence of a magnetic field further compresses the electron wave function. Let us first  define the cyclotron frequency and magnetic length induced by the components of an in-plane magnetic field $B_{x(y)}$ by,
\begin{align}\label{eq:lB}
   &\Omega_{x(y)}=\frac{eB_{x(y)}}{\sqrt{m_tm_l}},
   &l_{x(y)}=\sqrt{\frac{\hbar}{eB_{x(y)}}}.
\end{align}
We can then write,
\begin{align}\label{eq:om}
&\omega_{x}'=\omega_{x}\left(1+\frac{\Omega_{y}^2}{\omega_{x}^2}\right)^{1/2},
&\omega_{y}'=\omega_{y}\left(1+\frac{\Omega_{x}^2}{\omega_{y}^2}\right)^{1/2}. 
\end{align}
We note that the homogeneous magnetic field $\textbf{B}_0$ also has an out-of-plane component, $b_{0z}$. However, given the experimental setups for the integrated micromagnet, we have $b_{0z} \lesssim 0.02$ T \cite{Borjans19}. In this case, we find the confinement length due to $b_{0z}$ becomes $l_z \gtrsim 180$ nm which is far larger than other confinement lengths. As such, we can safely ingnore the out-of-plane component of $\textbf{B}_0$ and only consider the in-plane magnetic field.

\eref{eq:H_xyz_B} clearly gives rise to a separable envelope function in which the in-plane envelope functions are simply given by the harmonic-oscillator wave functions. The out-of-plane envelope functions are discussed in detail in \ocite{Hosseinkhani20}. While the excited states $\psi_{z,n}$ can be found from numerical calculations \cite{Hosseinkhani20}, we find the approximate solution for the ground state read,   
\begin{equation}
\label{eq:psi_0z}
  \psi_{z,0}(\tz) \simeq\\
   \frac{z_0^{-1/2}}{\Ai'(-r_0)} 
  \begin{cases}
      \Ai(-\te_{z,0})e^{-\frac{\Ai'(-\te_{z,0})}{\Ai(-\te_{z,0})}\tz}\: , & \tz > 0 \, \\
   \Ai(-\tz-\te_{z,0})\: . & \tz \leq 0\, 
  \end{cases}
\end{equation}
while the (normalized) ground state energy reads,
\begin{align}
\label{eq:E_0z}
\te_{z,0} \simeq r_0 -\tilde{U}_0^{-1/2}.
\end{align} 
Here Ai is the Airy function, Ai$'$ its first derivative, and $-r_0\simeq-2.338$ its  smallest root (in absolute value). We used here normalized position, $\tz=z/z_0$, energy, $\te_{z,0}=\e_{z,0}/\e_0$, and potential $\tu_0=U_0/\e_0$ for which the length and energy scales are given by,
\begin{align}
    z_0 &= \left[\frac{\hbar^2}{2m_leF_z}\right]^{1/3},&
    \e_0 = \frac{\hbar^2}{2m_lz_0^2}.
    \label{eq:e0}
\end{align}

The term $H_{||}$ in 
\eref{eq:Hc} is due to the couplings caused by the in-plane components of $\textbf{B}_0$  reading ($B_x=B_{||,x}+b_{0x}$ and $B_y=B_{||,y}$),
\begin{align}
\label{eq:H_||}
H_{||}&=-B_x\frac{e}{m_l}yp_z+B_y\frac{e}{m_l}xp_z -B_xB_y\frac{e^2}{m_l}xy.
\end{align}
Given the confinement Hamiltonian \eref{eq:Hc}, the valley-dependent envelope functions can be found by solving \eref{eq:VDenv}, as discussed in detail in \ocite{Hosseinkhani21}. The valley-coupling parameter $V_v(r)=V_v\mathcal{S}_\mathrm{int}(x,z)$ is also modeled in \eref{eq:VDenv} to be,
\begin{align}\label{eq:Vv}
    V_v=-i\mathcal{C}_0\frac{z_0\tu_0eF_z}{2k_0}\left(1-\left[1-\frac{1}{2\tu_0}+i\frac{k_0z_0}{\sqrt{\tu_0}}\right]^{-1}\right).
\end{align}
where $\mathcal{C}_0\simeq -0.2607$ originates from the  lattice-periodic parts of the Bloch function \cite{Hosseinkhani20,Hosseinkhani21}, and the interface function $\mathcal{S}_\mathrm{int}(x,z)$ is given by \eref{eq:Sint}.

\section{The valley-dependent envelope functions}\label{app:2}
Without going into details, we review from \ocite{Hosseinkhani21} the general solution for the valley-dependent envelope functions for the ground $(q=0)$ and excited $(q=1)$ valley-orbital states. One can find,
\begin{align}
    \label{eq:Psi_t_p} %
    \Psi_{xyz}^{\pm z,(q)}= \psi_{xyz,0} + \psi_{||} +  \psi_\mathrm{st}^{\pm z,(q)},
\end{align}
in which,
\begin{align}\label{eq:psi_par}
    \psi_{||}=&-iB_x\psi_{x,0}\psi_{y,1}\sum_{n=1}\alpha_n\psi_{z,n}\\
&+iB_y\psi_{x,1}\psi_{y,0}\sum_{n=1}\beta_n\psi_{z,n}
-B_xB_y\eta\psi_{x,1}\psi_{y,1}\psi_{z,0}\:,\nonumber
\end{align}
and 
\begin{align}\label{eq:psi_st}
    \psi_\mathrm{st}^{\pm z,(q)}=\psi_{y,0} \sum_{(m,n)\neq(0,0)} c_{m,n}^{ \pm z,(q)}\psi_{x,m}\psi_{z,n}.
\end{align}
The coefficients used above read (see \ocite{Hosseinkhani21} for details),
\begin{align}
\label{eq:alpha_n}
\alpha_n&=-\frac{1}{2}\hbar\frac{e}{m_l}\frac{y_0'}{z_0}\frac{\langle\psi_{z,0}|\partial/\partial \tilde{z}|\psi_{z,n}\rangle}{\e_{z,0}-\e_{z,n}-\hbar\w_y'},\\ \label{eq:beta_n}
\beta_n&=-\frac{1}{2}\hbar\frac{e}{m_l}\frac{x_0'}{z_0}\frac{\langle\psi_{z,0}|\partial/\partial \tilde{z}|\psi_{z,n}\rangle}{\e_{z,0}-\e_{z,n}-\hbar\w_x'},\\\label{eq:eta}
\eta&=-\frac{1}{4}\frac{e^2}{m_l}x_0'y_0'\frac{1}{\hbar\w_x'+\hbar\w_y'},\\
c_{m,n}^{+z,(q)}&=\frac{(-1)^q e^{-i\phi_v}\mathcal{F}_{m,n}-\mathcal{P}_{m,n}}{\e_{m,n}-\e_0},
\end{align}
where we defined,
\begin{align}\label{eq:FH_p}
    \mathcal{F}_{m,n}&=\int e^{-2ik_0z}\psi_{x,m}\psi_{z,n}H_s\psi_{x,0}\psi_{z,0}d^3r,\\
    \label{eq:PH_p}
    \mathcal{P}_{m,n}&=\int \psi_{x,m}\psi_{z,n}H_s\psi_{x,0}\psi_{z,0}d^3r,
\end{align}
and $H_s=V_v \mathcal{S}_\mathrm{int}(x,z) + U_{\mathrm{steps}}(x,z)$.

\end{document}